\documentclass[preprint]{revtex4-1} 

\usepackage{amsfonts} 
\usepackage{graphicx}
\usepackage{amsmath, amssymb}
\usepackage{slashed}
\usepackage{braket}
\usepackage{physics}
\usepackage[bottom]{footmisc}
\graphicspath{  {images/} }
\begin{document}


\title{An observer's perspective of the Unruh and Hawking effects - using coherent signals to extract information from a black hole}

\author{Satish Ramakrishna}
\email{ramakrishna@physics.rutgers.edu}
\affiliation{Department of Physics \& Astronomy, Rutgers, The State University of New Jersey, 136 Frelinghuysen Road
Piscataway, NJ 08854-8019}


\date{\today}

\vspace{40mm}

\begin{abstract}

The Unruh effect is one of the first calculations that explain what one would see when transiting between an inertial reference frame with its quantum field vacuum state and a non-inertial (specifically, uniformly accelerating) reference frame. The inertial reference frame's vacuum state would not correspond to the vacuum state of the non-inertial frame. That observer (in the non-inertial frame) would see radiation, with a corresponding Bose distribution and a temperature proportional to the acceleration.
In this paper, I compute the response of this non-inertial observer to a single frequency mode in the inertial frame and deduce that, indeed, the cumulative distribution (over the observer's proper time) of frequencies observed by the accelerating observer would be the Bose distribution with a temperature proportional to the acceleration. The conclusion is that the Unruh effect (and the related Hawking effect) is {\it generic}, in that it would appear with {\bf any} incoming incoherent state and the Bose distribution is obtained as a consequence of the non-inertial frame's motion, rather than some special property of the quantum vacuum.
As a consequence of the analysis of a coherent set of signals, I show to extract information from the spectrum that an accelerated observer would see (as well as from the radiation from a black hole).

\end{abstract}
\maketitle

\section{Introduction}

The Unruh effect refers to work done in 1976 by W. G. Unruh and others \cite{Unruh,Davies1,Davies2,Davies3,Davies4,Davies5,Davies6,Crispino} which examined the effect on particle detectors of an accelerated observer due to a scalar photon field in its ground state as observed by an inertial observer. The effect and its derivation from standard principles of quantum field theory are well-known and are, now, material for textbooks and on-line presentations about the effects of quantum field theory when applied to accelerated observers and curved space-times. There have been several proposals for how to detect the effect experimentally, some recent ones being\cite{Gooding, Soda}.

As an aside, after the first version of this paper was circulated, I have become aware of other work that have a similar approach\cite{Padma, Milonni, Rosabal}. The first two of these related papers start at the same initial point and proceed in other directions. This paper considers, as a difference to the other work, the Hawking effect as well as a straightforward experiment to verify the effects in the laboratory. In addition,  inspired by the response to a coherent set of signals,  I show how to extract information from the spectrum that an accelerated observer would see (as well as from the radiation from a black hole).

The difference between this paper and the approach in \cite{Padma, Milonni} is that while we all consider the ergodic realization of the Bose distribution seen by an accelerated observer from just one mode, this paper makes clear that any collection of incoherent, non-interfering modes will result in the same result. Essentially, one needs to take care not to send a collection of correlated modes with precise phase relationships between them - such a collection of modes would not produce a Bose distribution as seen by the accelerated observer. This would be true, for instance, if the modes were interacting and the quantum state were more complex. The quantum vacuum is a collection of completely incoherent oscillations - all the results obtained in the literature work in that case and only in other similar cases. This is explained in the section on extension to multiple modes. It is a novel result and has not been described in the literature. An explicit example of such a difference is shown in in the language of Bogolyubov transformations in the section at the middle of the paper.

We describe the usual approach\cite{Kempf}, working in one-dimension, with an inertial frame (coordinates $(t,x)$) and a frame uniformly accelerated in the positive-x direction with acceleration $a$. The coordinates in the accelerated frame are written as $(\tau,R)$. We use natural units i.e., $\hbar=1, c=1, G=1$ and the $+---$ metric signature, in all that follows. The accelerated observer is supposed to start traveling in the negative-x direction at the speed of light at $t = \tau = - \infty$ and end up traveling in the positive-x direction at the speed of light at  $t = \tau =  \infty$, with the position of the observer at $t = \tau = 0$ being $x = e^{a R}$. This path is displayed in Fig. 1 for two observers traveling with the same constant acceleration.

\begin{figure}[h!]
\caption{Trajectories of two equally spaced (in their frame) accelerated observers}
\centering
\includegraphics[scale=.75]{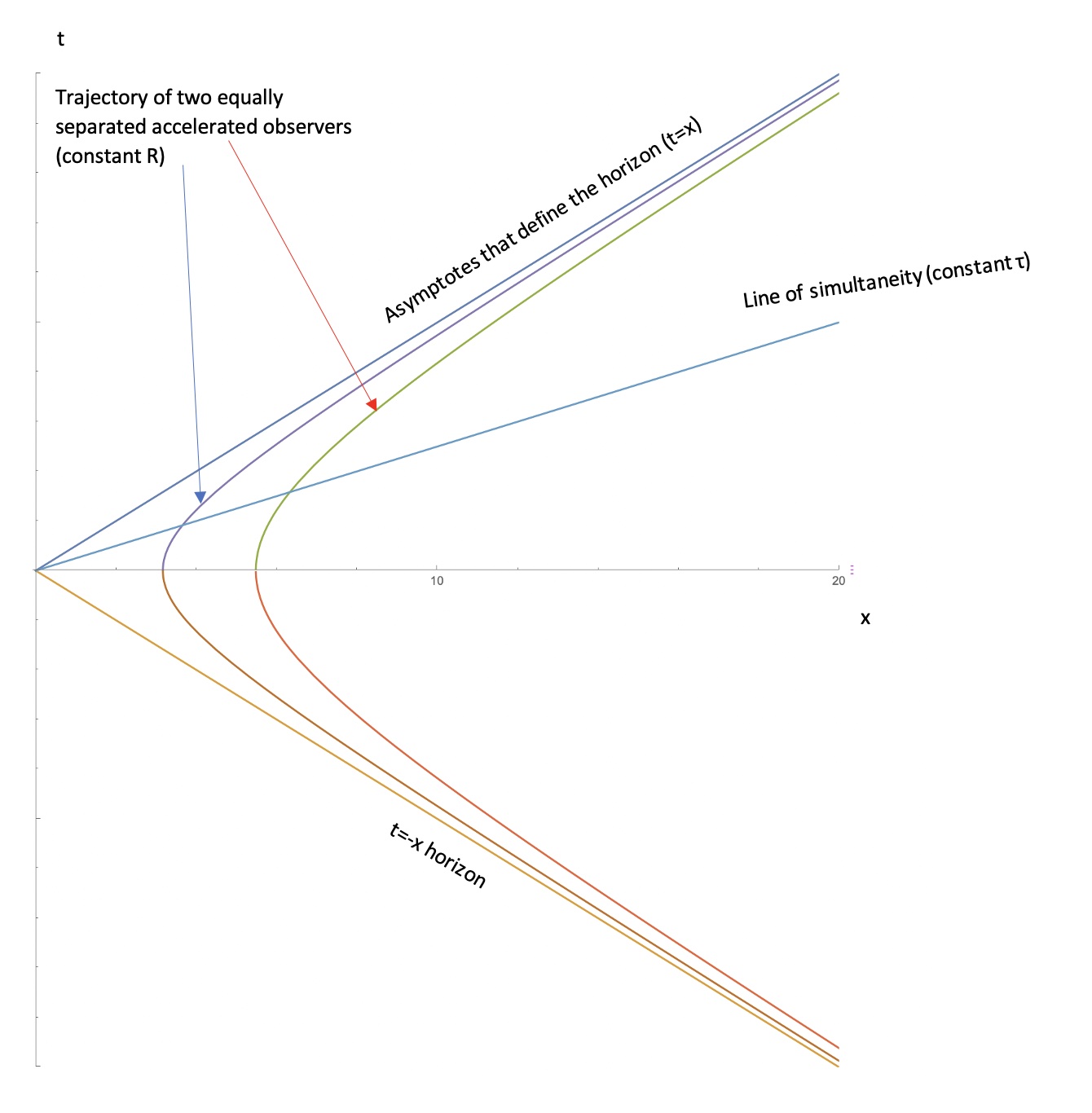}
\end{figure}

The analysis begins with writing variables for the inertial frame $(t, x)$ (usually, light-cone coordinates $u = t - x  \:, v = t + x$ are used)
while Rindler variables $(\tau, R)$ are used in the accelerated reference frame, with their light-cone coordinates ${\bar u} = \tau - R   \:, {\bar v} = \tau + R $.

The equations connecting these variables are standard \cite{Kempf}, i.e., 
\begin{eqnarray}
x = \frac{1}{a} e^{a R} \cosh(a \tau)  \: \: \: , \: \: \: 
t = \frac{1}{a} e^{a R} \sinh(a \tau)  \: \: \: , \: \: \: \nonumber \\
u = - \frac{1}{a} e^{a R} e^{-a \tau}  \: \: \: , \: \: \: 
v =  \frac{1}{a} e^{a R} e^{a \tau}   \: \: \:  \: \: \: 
\end{eqnarray}

The Rindler patch is restricted, in this case, to the domain $x >=0, -\infty <t<+\infty$ and the lines $t = \pm x$ are horizons that represent all that the accelerating observer will ever see of the inertial frame. In Fig.1, I show the constant-$R$ and constant-$\tau$ lines - the first are hyperbolas, while the second are straight lines passing through the origin. The horizon, in the Rindler coordinates, is described by $R=-\infty, \tau = \pm \infty$. The metrics in these spaces are, in the inertial, Minkowski space and in the accelerated observer's coordinates,
\begin{eqnarray}
(ds)^2 = (dt)^2 - (dx)^2 \: \: , \: \: 
(ds)^2 = e^{2 a R} \bigg( (d \tau)^2 - (dR)^2 \bigg) 
\end{eqnarray}
An accelerated observer standing at a point $R$ in their space would trace out a hyperbola (in the inertial coordinates), whose equation would be 
$t^2 - x^2 = \frac{1}{a^2} e^{2 a R}$.

To summarize, for a fixed $R$  (accelerated observer is stationary in their own frame), the `outside', i.e., inertial observer would see that observer's path as $x \propto \cosh (a \tau), t \propto \sinh(a \tau)$. This feature will be useful for us in the next couple of sections.

As an aside, there are many interesting phenomena connected to the structure of the Rindler coordinate system - one such is analyzed by Bell \cite{Bell}.

The next step is to write the scalar photon field $\phi$ in the two light-cone coordinates and establish a Bogolyubov transformation \cite{Unruh, Kempf} that relates the annihilation and creation operators in the two representations. That immediately implies that the number operator in the mode-expansion used by the accelerated observer will have a non-zero expectation value in the vacuum state of the inertial observer. This expectation turns out, for every frequency $\omega$ in the accelerated frame's mode-expansion to have the functional form $\sim \frac{1}{e^{\frac{2 \pi \omega}{a}}-1}$. This thermal radiation is seen by the accelerated observer (temperature $\frac{a}{2 \pi}$), despite the quantum field being in the vacuum state of the inertial observer.

\section{Motivation for an observer dependent approach}

In order to understand what an observer traveling along the trajectory $R=0$ would see, it is useful to think of what exactly is a fluctuation the observer could see and measure.

\subsection{What is a vacuum fluctuation}

It is a safe statement to state two requirements for a physical field fluctuation
\begin{enumerate}
\item a physical field fluctuation, i.e., not a vacuum fluctuation, is one that persists (at least through its antecedents or descendants) from $t = -\infty$ to $t = \infty$.
\item In addition, if the expectation of the number operator were calculated in the field state at a time-slice $t$, one should get a non-zero answer. 
\end{enumerate} 
In plain English, if we have a real particle in our reference frame, it came from some other particle(s) and disappeared, creating another particle(s), stretching back to $-\infty$ and to $\infty$ in time. If it simply emerged from the vacuum and disappeared into the vacuum, it would not be a physical particle, but would be called a vacuum fluctuation or a "virtual" particle.

For a vacuum fluctuation, one needs both of these requirements to be violated. The particle, or its antecedents and descendants should not persist between $t = -\infty$ and $t = \infty$ and at any time slice in that reference frame, the particle number should be zero.

This common-sense interpretation implies that a particle that exists between two times $t_1$ and $t_2$ in the inertial reference frame, with no antecedents or descendants is a virtual particle in that frame. The Feynman diagram for such a particle is shown in Fig. 2 (the irregular oval). However, as is clear, this particle, {\bf as far as the accelerated observer is concerned}, exists between $\tau=-\infty$ to $\tau=+\infty$. Such a particle would be considered a physical particle by the accelerated observer.

\begin{figure}[h!]
\caption{Dotted line - vacuum fluctuation in the inertial frame}
\centering
\includegraphics[scale=.75]{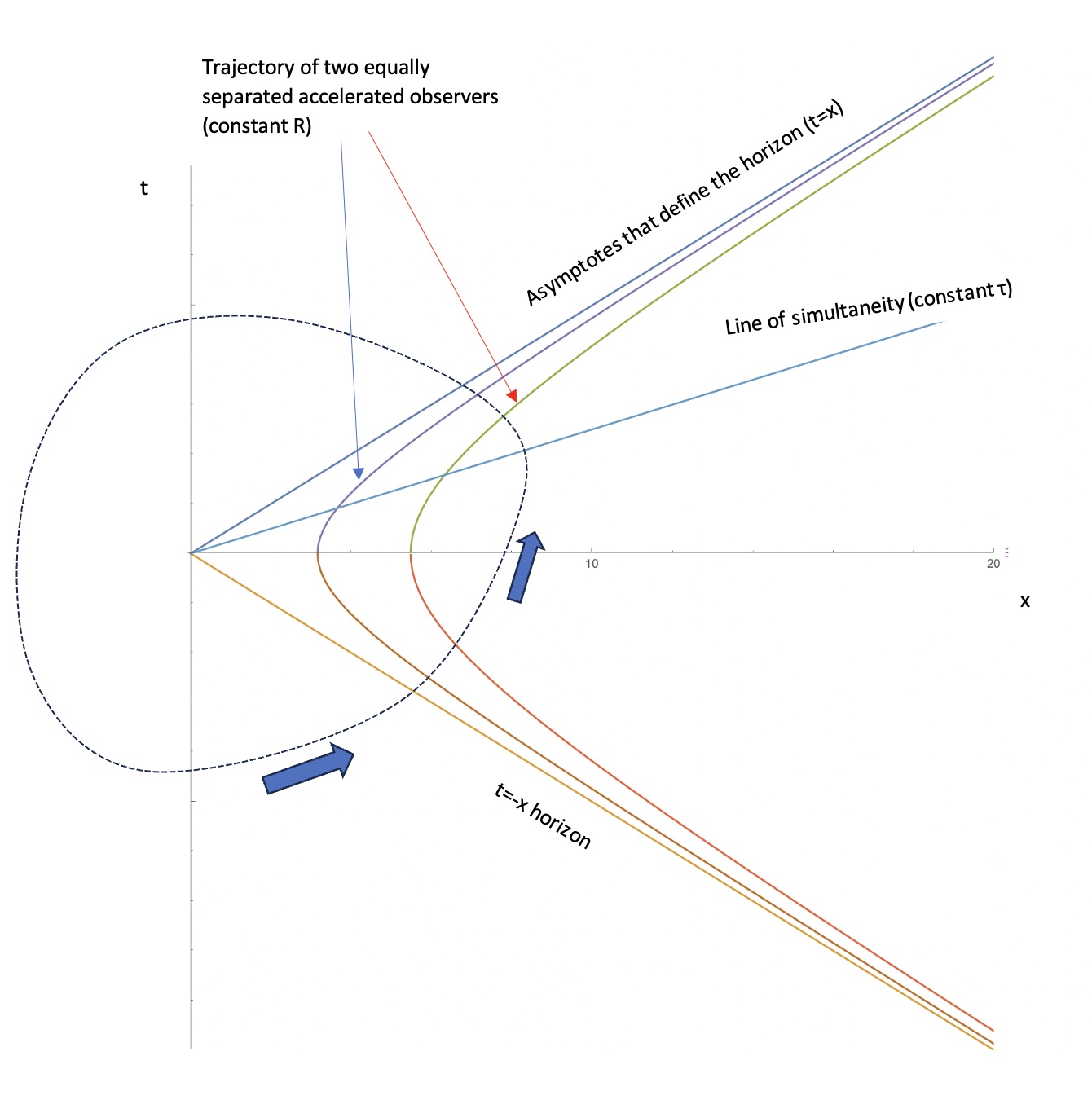}
\end{figure}

The vacuum fluctuation in the inertial frame meets the criteria to be a physical field fluctuation in the accelerated frame. Indeed, if one of the hyperbolic paths were that for the observer, the path of the field fluctuation would terminate at the detector and such a fluctuation would be observed. Depending upon the size of the vacuum fluctuation and exactly where it intersects the path of the accelerated observer, the observer would see different frequencies, since there is relativistic red-/blue-shift given the traveling speed of the observer. From the distribution of the energies of these vacuum fluctuations and their consequent persistence time, along with the appropriate red-/blue- shift, we could compute carefully the distribution of energies seen by the accelerated observer.

There is, however, a simple calculation one could do for one mode, with a fixed frequency. We start as follows : the speed of the observers can be computed from Equation (1), {\it viz.}  $v = \tanh(a \tau)$ in the proper time of the observer - interestingly, this is independent of $R$ which allows the calculation to proceed easily. 

Let's now look at the effect of sending a single physical mode, frequency $\omega_0$  from somewhere in the inertial frame, towards the accelerated observer.

\subsection{The effect on the accelerated observer}

The accelerated observer would see the frequency of the mode as 
$ \omega_{obs} = \omega_0 \sqrt{\frac{1-v}{1+v}}$, which is the usual relativistic red-shift formula. $v$ is the speed of the observer away from or towards the mode's propagation direction (we are working in 1-dimension here). In this case, since $v = \tanh(a \tau)$, we deduce that $ \omega_{obs} = \omega_0 e^{-a \tau}$.

If the accelerated observer were to count wavefronts, i.e., the accumulated phase over some period of proper time, one already has an answer for this quantity from the inertial frame. In fact, this quantity should be invariant regardless of which frame it is calculated in. It should be, over a time interval $t_0$ to $t$, equal to $\omega_0 \times (t - t_0)$. This can be written in the coordinates of the accelerating observer as $\omega_0 \int_{\tau_0}^{\tau} ds \: e^{-a s}$. 
Hence, the single (positive frequency) mode would be received as 
$f(\omega_0)=N_0 e^{-i \omega_0 \int_{\tau_0}^{\tau} ds \: e^{-a s}}$
where we need to apply a UV cut-off for how closely the accelerated observer can get to the horizon. $N_0$ is a normalization factor and is usually chosen to be $\frac{1}{\sqrt{\omega_0}}$.

Another way of stating the above is that the waves in the inertial frame are functions of $t \pm x$, which, when written in the accelerated observer's coordinates has the form shown.

Note, also, that if the incoming pulse were not a pure mode, but actually the complex conjugate of $f(\omega_0)$, the received signal would be a pure mode.

This can be simplified to (where ${\cal C}$ is a phase) to ${\cal C} N_0 e^{i \frac{\omega_0}{a} e^{-a \tau}}$
which is a chirp-type signal (see Fig. 3).

\begin{figure}[h!]
\caption{Chirp Signal Example}
\centering
\includegraphics[scale=.75]{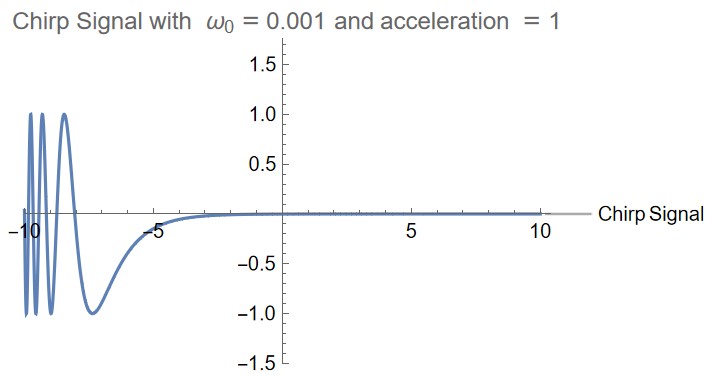}
\end{figure}

The next, and novel step, is to compute the Fourier transform of this frequency-modulated signal\cite{note}. We can compute the positive (${\cal F}(\Omega)$) and negative (${\cal G}(-\Omega)$) frequency components as below (we treat $\Omega$ as positive),

\begin{eqnarray}
{\cal F}(\Omega) = {\cal C} N_0 \int_{-\infty}^{\infty} d \tau \frac{1}{N_{\Omega}} e^{i \Omega \tau} \:  e^{i \frac{\omega_0}{a} e^{-a \tau}} \nonumber \\
{\cal G}(-\Omega) = {\cal C} N_0 \int_{-\infty}^{\infty} d \tau \frac{1}{N_{\Omega}} e^{- i \Omega \tau} \:  e^{i \frac{\omega_0}{a} e^{-a \tau}}
\end{eqnarray}
Again, here, $N_{\Omega}$ is a normalization factor for the real-space function and is chosen to be $\frac{1}{\sqrt{\Omega}}$. The integrals can be computed using standard methods\cite{Kempf} -
\begin{eqnarray}
{\cal F}(\Omega) = {\cal C} \frac{N_0}{N_{\Omega}}  \:  \frac{1}{a} \: e^{\frac{ \pi \Omega}{2 a}} \: e^{i \frac{\Omega}{a} \ln(\frac{\omega_0}{a})} \Gamma(\frac{-i \Omega}{a}) \nonumber \\
{\cal G}(-\Omega) = {\cal C} \frac{N_0}{N_{\Omega}}  \:  \frac{1}{a} \: e^{\frac{- \pi \Omega}{2 a}} \: e^{-i \frac{\Omega}{a} \ln(\frac{\omega_0}{a})} \Gamma(\frac{i \Omega}{a})
\end{eqnarray}

so that 
\begin{eqnarray}
|{\cal F}(\Omega)|^2 = \frac{\frac{2 \pi}{a \omega_0}}{e^{\frac{-2 \pi \Omega}{a}}-1} \: \: , \: \: 
|{\cal G}(-\Omega)|^2 = \frac{\frac{2 \pi}{a \omega_0}}{e^{\frac{2 \pi \Omega}{a}}-1} 
\end{eqnarray}

which is obtained for {\it just one emitted mode} and represents the central result of the paper. One does not need to deal with a vacuum state with a myriad of modes to see a thermal distribution - this is just the result of the particular frequency modulation one gets with an accelerated observer. With one mode, the detector would need to be analyzed after its entire journey is complete to get a complete picture of the Bose distribution. The result is generic and is related precisely to the existence of the horizon and would be observed, as discussed below, to any {\it {incoherent}} set of waves sent towards the accelerated observer.

To reiterate, a pure emitted mode can produce a thermal distribution of received radiation purely due to kinematic transformation to the accelerated observer's reference frame. This and the conclusions drawn in the next section has implications for other interesting problems in physics, for instance, the black-hole information problem\cite{Hawking2}.

\subsection{Extension to multiple modes}

If the inertial frame sent in a collection of {\underline {incoherent}} modes with a distribution $f(\omega)$ towards the accelerated observer, the consequent radiation received would be just a sum over the respective $|{\cal F}|^2$ or $|{\cal G}|^2$ for each mode, so we recover exactly the same distribution for each, times a pre-factor.  In detail, this is
\begin{eqnarray}
\Sigma = \int d\Omega \: {\cal F}(\Omega)
\end{eqnarray}
and we would be interested in $|\Sigma|^2$ as a function of $a$. But it is obvious that unless cross terms in the below integral average out to $0$, the distribution that results would be exceptionally complicated, certainly not the Bose distribution.
\begin{eqnarray}
|\Sigma|^2 = \int d\Omega \: \int d\Omega^{'} \: {\cal F}(\Omega) {\cal F}^{*}(\Omega^{'}) \ne  \int d\Omega \: |{\cal F}(\Omega)|^2
\end{eqnarray}
Vacuum fluctuations are uncorrelated across space and time and hence completely incoherent. Since individual modes are uncorrelated with each other, one obtains the usual Unruh result (also the Hawking result, as explained in the next section). If, however, the modes were correlated, it is possible for information to be stored in the phase relationships between the modes, that would determine (and possibly be read out of) the particular distribution that results. This is demonstrated in the next section.

In conclusion, if a coherent set of waves were sent, there are interference terms, related to the phase-factors in Equations (6) and (7), which complicate the picture and prevent the result from being a pure sum over the Bose distribution.

\section{An explanation in the Bogolyubov language}

\begin{figure}[h!]
\caption{Pair Creation - near the horizon}
\centering
\includegraphics[scale=.75]{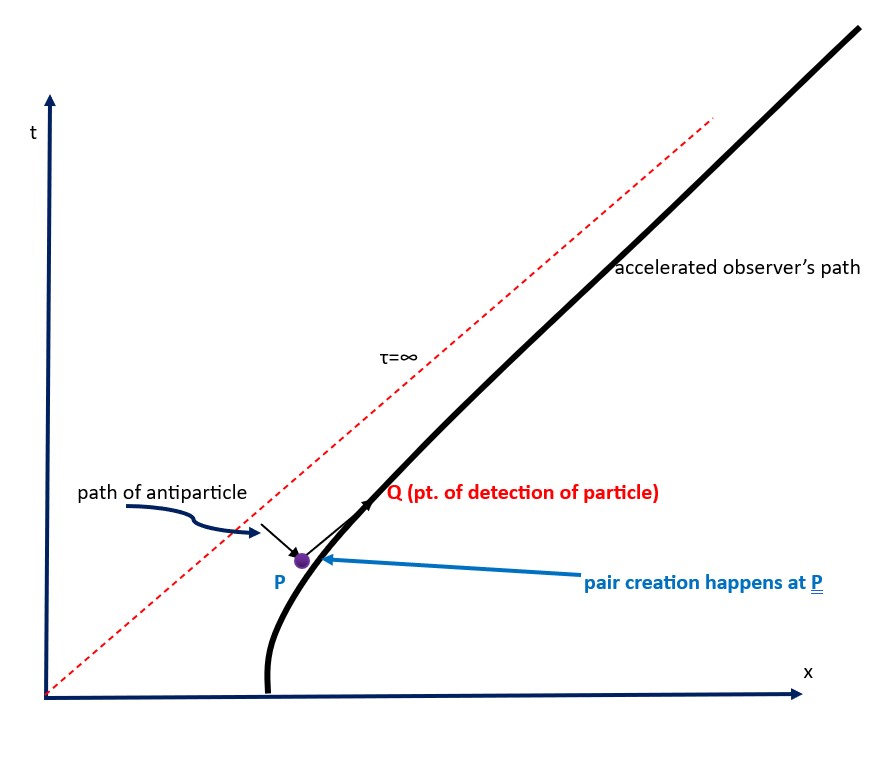}
\end{figure}

Suppose the inertial observer initially prepares the universe in the vacuum state of the scalar field: we work in one-dimension. In this inertial observer's frame, there are vacuum fluctuations (see Fig. 4).  Next, suppose a particle anti-particle pair is produced near the accelerated observer's horizon, at point P in the figure. One particle stays within the accelerated observer's horizon and the other heads to the horizon to cross it, though crossing it will take an infinite amount of time in the accelerated observer's reference frame. Once the accelerated observer captures the particle that crosses their path (at point Q), the partner particle is now single and alone in the inertial observer's frame. It is possible, in fact, necessary, that the accelerated observer's detector will eventually emit the energy it has thus absorbed, however for some time in the inertial observer's frame, the scalar field is no longer in the vacuum state - it has one free particle (the other has been absorbed). This can continue for more particles - the resulting states are neither precisely the vacuum state (there are free particles), nor the one- (or more-) particle states (these particles have no antecedents in the inertial frame). Hence, one should expect, as more vacuum fluctuations occur, that the eventual state, in the frame of the inertial observer can be written as a {\bf superposition} of various inertial frame Fock space states.

Depending on the exact frequency and wave-vector of particles absorbed by the detector, the state of the scalar field evolves from a simple vacuum state (initially) to a general superposition, with $\vec n$ being the number of particles in all the possible energy states (in the inertial observer's basis)
\begin{eqnarray}
|\psi> = \sum_{\vec n=0}^{\infty} \:C_{\vec n}\: |\vec n>
\end{eqnarray}

In this general state, the accelerated observer will compute some distribution of radiation. That distribution, as we will see, will not be the Bose distribution - though initially it would start out close to one. {\bf Is it possible for the accelerated observer to reconstruct the state from observing the spectrum of the radiation?} We will answer this question below. Let's set up the problem first.

In the usual method of deriving the Unruh effect, as for instance in \cite{Kempf}, one ascribes the effect as due to fact that field creation/annihilation operators for the massless scalar field, written for the inertial and accelerated observers are related by a Bogolyubov transformation. Essentially, the vacuum state seen by the two observers is different. If the inertial observer sets the universe up in a vacuum state from their point of view (the $u, v$ coordinates in Equation (1)), with their creation/annihilation operator acting on it as usual (we work, as mentioned, in one dimension, and for a massless scalar field, frequency $\omega$ equals the wavevector $k$ in natural units)
\begin{eqnarray}
a(\omega) |0> = 0
\end{eqnarray}
the accelerated observer, with coordinate system $\bar u, \bar v$ writes their creation/annihilation operators \cite{Kempf} as 
\begin{eqnarray}
b({\Omega}) =\int_{0}^{\infty} \: d \omega \bigg( \alpha_{\Omega \omega} a(\omega) - \beta_{\Omega \omega} a^{\dagger}(\omega)  \bigg)
\end{eqnarray}
and by comparing mode expansions in the common region of the two reference frames, we get
\begin{eqnarray}
 \alpha_{\Omega \omega} = + \frac{1}{\sqrt{2 \pi a}} \sqrt{\frac{\Omega}{\omega}} \:  e^{+ \frac{\pi \Omega}{2a}} \:  e^{i \frac{\Omega}{a} \ln{\frac{\omega}{a}}} \: \Gamma(-i \frac{\Omega}{a}) \nonumber \\
 \beta_{\Omega \omega} = - \frac{1}{\sqrt{2 \pi a}} \sqrt{\frac{\Omega}{\omega}} \:  e^{- \frac{\pi \Omega}{2a}} \:  e^{i \frac{\Omega}{a} \ln{\frac{\omega}{a}}} \: \Gamma(-i \frac{\Omega}{a}) 
\end{eqnarray}

The number operator in the accelerated observer's frame is $N_b = b^{\dagger}(\Omega) b(\Omega)$ and can be written as
\begin{eqnarray}
b^{\dagger}(\Omega) b(\Omega) = \int_{0}^{\infty} \: d \omega \: \int_{0}^{\infty} \: d \omega^{'} \: \: \: \: \: \: \: \: \: \: \: \: \: \: \: \: \: \: \: \: \nonumber \\
\bigg( \alpha^{*}_{\Omega \omega} a^{\dagger}(\omega) - \beta^{*}_{\Omega \omega} a(\omega) \bigg) \bigg( \alpha_{\Omega \omega^{'}} a(\omega^{'}) - \beta_{\Omega \omega^{'}} a^{\dagger}(\omega^{'})  \bigg)
\end{eqnarray}

Now, in the later evolved state $|\psi>$, the expectation value of the number operator can be written as
\begin{eqnarray}
<\psi|b^{\dagger}(\Omega) b(\Omega)|\psi> = \int_{0}^{\infty} \: d \omega \: \int_{0}^{\infty} \: d \omega^{'} \:  \sum_{\vec m,\vec n=0}^{\infty} \nonumber \\
C_{\vec m}^{*} C_{\vec n} \: \bigg(  \alpha^{*}_{\Omega \omega} \alpha_{\Omega \omega^{'}} \delta_{\vec m; n_{\omega}+1, n_{\omega^{'}}-1} \sqrt{n_{\omega^{'}}} \sqrt{n_{\omega}+1}\nonumber \\
+ \beta^{*}_{\Omega \omega}  \beta_{\Omega \omega^{'}}  \delta_{\vec m; n_{\omega}-1, n_{\omega^{'}}+1} \sqrt{n_{\omega^{'}}+1} \sqrt{n_{\omega}} \nonumber \\
- \alpha^{*}_{\Omega \omega} \beta_{\Omega \omega^{'}} \delta_{\vec m; n_{\omega}+1, n_{\omega^{'}}+1} \sqrt{n_{\omega^{'}}+1} \sqrt{n_{\omega}+1}  \nonumber \\
-\beta^{*}_{\Omega \omega} \alpha_{\Omega \omega^{'}}  \delta_{\vec m; n_{\omega}-1, n_{\omega^{'}}-1} \sqrt{n_{\omega^{'}}} \sqrt{n_{\omega}}
 \bigg)
\end{eqnarray}
To get a properly regularized answer, one needs to add regularization terms to the integrals over $\omega$ and $\omega^{'}$. Then, inspecting the form of $\alpha$ and $\beta$, we find that the expectation value of the number operator, i.e., the spectrum of the radiation in the frame of the accelerated observer can be written as an infinite series of powers of exponentials of the form $e^{-2 \pi \frac{\Omega}{a}}$. These are a proper set of independent basis functions, so that the coefficients can be determined from an inspection of the spectrum.

As an example, if we restrict ourselves ourselves only to one-particle states (denoted by $|1_i>$ for the one-particle state with energy $\omega_i$) and the vacuum (denoted by $|0>$), we could write
\begin{eqnarray}
|\psi> = C_0 |0>+\sum_{i=1}^{\infty} \:C_{i}\: |1_i>
\end{eqnarray}
Carrying through the calculation,  as in Equation (13), from the terms sandwiched by the $<0|$ and $|0>$ states, we will get the usual Bose distribution result\cite{Kempf}. Note that the terms in Equation (12) with two creation (or annihilation) operators yield $0$, but we will have cross-terms with $<j|$ and $|i>$ sandwiching one creation operator for the state $<j|$ and one annihilation operator for the state $|i>$. Those immediately give non-Bose distribution terms in the computation.  However, we now have a sequence of powers of $e^{-2 \pi \frac{\Omega}{a}}$ multiplied by products $C_i^{*} C_j$. Since these exponentials form a linearly independent basis, we can deduce, from the resultant distribution, several equations which can be solved to yield the complex coefficients $C_i$. Hence, we can deduce what vacuum fluctuations were created by analyzing the non-Bose distribution produced - this has relevance to the information paradox and explicitly shows that information can be recovered from the distribution observed by the accelerated observer.

The argument goes through in similar fashion for more varieties of states- with two and more particles. One can hence deduce the state of the system from quantitatively measuring components of the observed (non-Bose) particle distribution.

\section{Extension to Hawking's calculation}

Hawking radiation is obtained through a similar calculation, with a non-flat event-horizon\cite{Kempf, Hawking1}. The same considerations apply here, i.e., it is generic and doesn't apply to just the vacuum state, but to any general state with an incoherent set of modes. Let us study this in detail.

We study the situation in two coordinate systems - the familiar Schwarzschild and Kruskal-Szekeres (KS) and write the regular Schwarzschild metric (with $r_S=2 M$)
\begin{eqnarray}
(ds)^2 = (1 - \frac{r_S}{r}) (dt)^2 - \frac{(dr)^2}{(1 - \frac{r_S}{r})} - r^2 (d \Omega)^2
\end{eqnarray}
and the transition to the KS metric is achieved with (for $r >= r_S$)
\begin{eqnarray}
FOR:  r >= r_S  \: \: 
T = \sqrt{\bigg(\frac{r}{r_S}-1 \bigg)} \: \:  e^{\frac{r}{2 r_S}} \: \: \sinh \frac{t}{2 r_S}  \nonumber \\
X = \sqrt{\bigg(\frac{r}{r_S}-1 \bigg)} \: \:  e^{\frac{r}{2 r_S}} \: \: \cosh \frac{t}{2 r_S}
\end{eqnarray}

and the usual Schwarzschild metric in these coordinates is $(ds)^2 = \frac{32 M^3}{r} e^{\frac{r}{r_S}} \: \bigg( (dT)^2 - (dX)^2 \bigg) + r^2 (d \Omega)^2$.

We see that for fixed $r$, the Hawking effect computation is analogous to that for the Unruh effect, except for a visible difference in Equation (16), where for $r > r_S$, there is a pre-factor $\sqrt{r/r_S-1}$, which is sub-dominant compared to the exponential in $r$ for large $r$. There is also a difference in how the $r$ space is mapped - the line $T=X$ is now $r=r_S$, not $r = -\infty$ as in the Unruh example. However, in the limit of large $r$ ($r \rightarrow \infty$), this is irrelevant. 

Explicitly: In the Unruh effect,  the space of $(t,x)$ is Minkowski, while  $(\tau,R)$ is the Rindler space. A pure mode in Minkowski leads to a chirp in Rindler (as was shown in Section II), hence produces a thermal distribution in the Rindler space, which observers infinitely far away would realize was a consequence of their acceleration $a$. The temperature they would compute from the distribution would be $\frac{a}{2 \pi}$. Note that to do this, the observers would need to collect particles and deduce the temperature from the distribution obtained cumulatively when they got infinitely far away. 

For the Hawking effect,  $(T,X)$ is Minkowski for fixed $r$. And $(t,r)$, the real space, is Schwarzschild, which appears related to the $(T,X)$ space by a Rindler-type transformation, an approximation that gets better for large $r$. Identically, a pure mode in Minkowski produces a chirp in real space (quasi-Rindler for large $r$), which leads to a {\underline {thermal distribution}} in this Rindler space (where we look in the large $r$ region). And, to re-formulate the argument in Section II, $X \propto \cosh ({\cal A} t), T \propto \sinh({\cal A} t)$, where ${\cal A} = \frac{1}{2 r_S}$. Hence, in the $(t,r)$ space, which is the coordinate system used by observers in real space, the quantity corresponding to the acceleration (as in the calculation for the Unruh effect) would be ${\cal A} =\frac{1}{2 r_S}$ (again, for large $r$). This leads to a temperature that observers at infinity would observe, of $\frac{1}{4 \pi r_S} = \frac{1}{8 \pi M}$, i.e., the usual Hawking result.

If the signal in the Minkowski space were not a collection of non-interacting pure notes (exactly as argued in IIC), the distribution would not be the Bose distribution, arguing similarly to the previous case. Essentially, if the black hole were immersed in such an entangled state, similarly, the radiation emitted would {\it not} be Bose distributed.  Again, the distribution would start out as the Bose distribution and would evolve from there. This is a novel realization that does not appear to have made it to the literature. 

\section{Discussion}

The key points in the argument that the chirp observed by an accelerated observer, which is a frequency set from $0$ to $\infty$ in spread and of the particular functional form obtained in Section II, is required for the Bose distribution to be realized. 
This discussion has implications for how to experimentally verify the Unruh and Hawking effect. One does not need to locate a suitable black-hole or look for vacuum fluctuations. One can simply shoot a pure mode at an accelerated detector, as in Fig. 5 and verify that the spectrum obtained, when averaged over time, is the Bose distribution. A schematic of the experiment is indicated - positively charged multi-level ions (set up to be in a particular intermediate energy level) are shot at high speed (close to the velocity of light) at a source of a single frequency. An electric field decelerates, then accelerates the molecules in the opposite direction to a high speed (again as close to the velocity of light as possible). Once an equilibrium situation is set up, the field (and frequency source) is turned off - the molecules drift at constant speed towards either end of the apparatus where they are classified by the energy they deliver to the targets at the opposite ends (deduced from spikes in electron current flow as they de-ionize).  One then obtains the statistics of what wavelength was absorbed at each point on the path - essentially one has $\omega_{obs}$ at several points on the path. A discrete Fourier transform can now be calculated and compared to the expected distribution - in this case, we should obtain the Bose distribution. 

Interestingly, in this situation, the molecules have a constant acceleration in their rest frame owing to the transformation properties of the longitudinal component (along the direction of travel) of the electric field\cite{Jackson}. Basically,  if $E_{horizontal}$ is the electric field along the path of the ions, then the ions would see the same field in their respective rest frames, so their acceleration would be uniform in their rest frame,.

We would essentially be verifying the relativistic red/blue-shift phenomenon and its consequent Bose spectrum, thereby verifying the Unruh and Hawking effects in a quantum vacuum.

We could also verify a non-Bose distribution is obtained if we sent a coherent set of plane waves of different frequency, but with definite phase relationship, from the source of radiation. The resulting ergodic average of radiation received would {\it {not}} be the Bose distribution, as discussed in Section IIC. The actual distribution would be obtained by computing the actual product of integrals in Equation (7).

\begin{figure}[h!]
\caption{An experimental schematic}
\centering
\includegraphics[scale=.45]{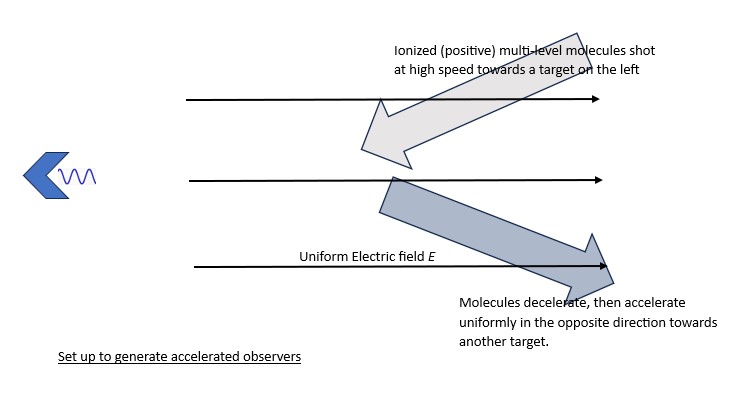}
\end{figure}

\section{Conclusions}

We have shown explicitly that Unruh (and Hawking) radiation is generic and is obtained as a general feature of a non-inertial, uniformly accelerated reference frame. In addition, the thermal radiation could well emerge from pure modes in a different reference frame. However, if the accelerated observer (in the Unruh case) is immersed in an entangled quantum field state, the distribution of radiation observed would not be the Bose distribution. The same happens for particle emission at the black hole horizon. It is possible to deduce the quantum state from an analysis of the spectrum of the observed radiation.

Indeed, the experiments devoted to checking for the Unruh and Hawking effects could well use a single mode, replace the ensemble average by an ergodic average and verify these rather straightforward results with the simple set-up described in the paper.

\acknowledgments
Detailed discussions and helpful criticism from Scott Thomas were especially useful and I thank him for this and other rather educational excursions. Dr. Achim Kempf's rather clear discussions were inspiring. The hospitality of the Rutgers Physics Department and the NHETC is also gratefully acknowledged. A very careful and exhaustive review by an anonymous referee is greatly appreciated.


\begin{thebibliography}{10}
\bibitem{Unruh} Unruh, W. G. \textit{Notes on black-hole evaporation} Phys. Rev. D {\bf {14}}, 4  (1976)
\bibitem{Davies1} Davies, P. C. W.,  \textit{Scalar particle production in Schwarzschild and Rindler metrics}  J. Phys. A 8, 609–616 (1975)
\bibitem{Davies2} Davies, P. C. W., \textit{On the origin of black hole evaporation radiation}  Proc. Roy. Soc. London Ser. A 351, 129–139 (1976)
\bibitem{Davies3} Davies, P. C. W., Dray, T. , \& Manogue,  C. A.  \textit{Detecting the rotating quantum vacuum} Phys. Rev. D 53, 4382–4387 (1996)
\bibitem{Davies4} Davies, P. C.W., \&  Fulling, S. A.  \textit{Quantum vacuum energy in two dimensional space-times} Proc. Roy. Soc. London Ser. A 354, 59–77 (1977)
\bibitem{Davies5} Davies, P. C. W., \& Fulling, S. A.  \textit{Radiation from moving mirrors and from black holes} Proc. R. Soc. Lond. Ser. A 356, 237–257 (1977)
\bibitem{Davies6} Davies, P. C. W., Fulling, S. A. , \&  Unruh, W. G.  \textit{Energy-momentum tensor near an evaporating black hole} Phys. Rev. D 13, 2720–2723 (1976)
\bibitem{Crispino} Crispino, L., Higuchi, A., Matsas, G. \textit{The Unruh effect and its applications}  	Rev.Mod.Phys. 80:787-838 (2008)
\bibitem{Gooding} Gooding, C., Biermann, S., Erne, S., Louko, J., Unruh, W. G., Schmiedmayer, J., \& Weinfurtner, S. \textit{Interferometric Unruh Detectors for Bose-Einstein Condensates}Phys. Rev. Lett. 125, 213603 (2020)
\bibitem{Soda} Šoda, B., Sudhir, V., \& Kempf, A. \textit{Acceleration-Induced Effects in Stimulated Light-Matter Interactions} Phys. Rev. Lett. 128, 163603 (2022)
\bibitem{Padma} Srinivasan, K. , Sriramkumar, L. and Padmanabhan, T. \textit{Plane waves viewed from an accelerated frame: Quantum physics in a classical setting} Phys. Rev. D 56, 6692-6694 (1997) 
\bibitem{Milonni} Alsing, P., Milonni, P. \textit{Simplified derivation of the Hawking-Unruh temperature for an accelerated observer in vacuum} Am.J.Phys. 72 (2004) 1524-1529
\bibitem{Rosabal} Rosabal, J. A. \textit{New Perspective On The Unruh Effect} Phys. Rev. D 98, 056015 (2018)
\bibitem{Kempf} A rather good introduction is by Achim Kempf, available at https://uwaterloo.ca/physics-of-information-lab/teaching/quantum-field-theory-cosmology-amath872phys785-w2022
\bibitem{Bell} Bell, J. S. \textit{ How to teach special relativity}  Progress in Scientific culture 1(2) (1976), pp. 1–13. Reprinted in J. S. Bell: \textit{Speakable and unspeakable in quantum mechanics} (Cambridge University Press, 1987), chapter 9, pp. 67–80.
\bibitem{Hawking1} Hawking, S. W. \textit{Particle Creation by Black Holes} Commun. Math. Phys. 43 (1975) 199; Erratum: Commun. Math. Phys. 46 (1976) 206.
\bibitem{note}  The $a \rightarrow 0$ limit has to be taken right at the start, when we calculate $\omega_{obs}$, since for $a=0$, there are no horizons - the phenomenon is not continuously connected to the non-zero $a$ case.
\bibitem{Hawking2} Hawking, S. W. \textit{The Information Paradox for Black Holes} arXiv:1509.01147 [hep-th]
\bibitem{Jackson} Jackson, J. D. \textit{Classical Electrodynamics} 3rd edition, Wiley ISBN: 978-0-471-30932-1
\end{thebibliography}
\end{document}